\documentclass[aps,twocolumn,showpacs]{revtex4}
\usepackage{graphicx}
\usepackage{amsfonts}
\usepackage{amssymb}
\usepackage{amsbsy}
\usepackage{amsmath}
\usepackage{mathrsfs}
\usepackage{latexsym}
\usepackage{natbib}
\usepackage{bm}
\usepackage{color}
\usepackage{braket}
\usepackage{slashed}
\usepackage[normalem]{ulem}

\DeclareMathOperator{\arcsinh}{asinh}

\def\x{\mathbf{x}}

\def\k{\mathrm{k}}

\def\mr{\mathcal{M}}

\begin{document}

\author{Golam Mortuza Hossain}
\email{ghossain@iiserkol.ac.in}

\author{Susobhan Mandal}
\email{sm17rs045@iiserkol.ac.in}

\affiliation{ Department of Physical Sciences, 
Indian Institute of Science Education and Research Kolkata,
Mohanpur - 741 246, WB, India }
 

\date{\today}

\title{Equation of states in the curved spacetime of slowly rotating 
degenerate stars}

\begin{abstract}
We compute the equation of state for an ensemble of degenerate fermions by 
using the curved spacetime of a slowly rotating axially symmetric star. We show 
that the equation of state computed in such curved spacetime depends on the 
gravitational time dilation as well as on the dragging of inertial frames, 
unlike an equation of state computed in a globally flat spacetime. The effect 
of gravitational time dilation leads to a significant enhancement of the 
maximum mass limit of a degenerate neutron star. However, such an enhancement 
due to the frame-dragging effect is extremely small. Nevertheless, in general 
relativity the frame-dragging effect is crucial for computing angular momentum 
of the star which is also shown to be enhanced significantly due to the usage of 
curved spacetime in computing the equation of state. 
\end{abstract}

\maketitle

\section{Introduction}

The unprecedented observation involving both gravitational waves as well 
as electromagnetic waves originating from merger of a binary neutron star 
system \cite{Abbott_2017} has opened up a new window to probe quantum dynamics 
of matter fields in a spacetime where gravity is very strong. With this dawn of 
multi-messenger astronomy \cite{margalit2017constraining, radice2018gw170817, 
annala2018gravitational, branchesi2016multi, meszaros2019multi}, it should now
be possible to test various aspects of quantum field theory in curved spacetime 
with an accuracy never achieved before. However, the equation of states (EOS)
that are used to describe degenerate nuclear matter of these stars, are often
computed in a globally flat spacetime. We refer to such equation of states as 
the `flat EOS' \cite{shen2002complete, douchin2001unified, 
lattimer2016equation, tolos2016equation, ozel2016dense, katayama2012equation}.
Recently, the equation of states by using the \emph{curved} spacetime of 
\emph{spherical} stars have been computed for an ensemble of non-interacting 
degenerate fermions \cite{hossain2021equation}, as well as for interacting 
degenerate fermions within the so-called $\sigma-\omega$ model of nuclear 
matter \cite{hossain2021higher}. These equation of states computed in the 
curved spacetime, henceforth referred to as the `curved EOS', incorporate the 
effect of gravitational time dilation, unlike the flat EOS. However, 
astrophysical stars are spinning objects. So the spacetime geometry of such 
stars are better described by an \emph{axially} symmetric metric rather than a 
metric having spherical symmetry.

In this article, we present a first-principle derivation of the equation of 
state for an ensemble of degenerate fermions by using the curved spacetime of a
slowly rotating, axially symmetric star. The equations governing the spacetime 
metric of such a slowly rotating star can be studied through the approach of 
Hartle and Thorne \cite{hartle1967slowly, hartle1968slowly}. In this approach, 
an additional non-trivial component of Einstein's equation arises apart from
the set of equations, known as the Tolman-Oppenheimer-Volkoff (TOV) equations 
\cite{tolman1939static, oppenheimer1939massive} that governs the interior 
metric of a spherical star. For a given matter EOS, these equations are then 
solved to find observationally relevant properties such as the mass-radius 
relations of these stars \cite{baumgarte1999maximum, lyford2003effects}.

In order to compute the curved EOS here we use the methods of thermal quantum 
field theory \cite{laine2016basics, kapusta1989finite, das1997finite} in the 
curved spacetime \cite{hossain2021equation, hossain2021higher}, unlike other 
approaches where one uses Minkowski spacetime for analogous computation 
\cite{cook1994rapidly, cookrapidly}. We show that the EOS computed in the 
curved spacetime of a slowly rotating star depends on the gravitational 
time dilation as well as on the \emph{dragging} of inertial frames 
\cite{wex1999frame, ciufolini2004confirmation, cui1997evidence}. Both of these 
effects lead to a relatively \emph{stiffer} EOS. The effect of gravitational 
time dilation however is much stronger compared to the effect of frame-dragging. 
The resultant maximum mass limit as implied by the curved EOS is shown to be 
significantly higher than the one implied by the corresponding flat EOS.

\section{Fermions in curved spacetime}

In this section we briefly review Fock-Weyl formulation of Dirac action that 
governs the dynamics of a free fermion in curved spacetime. In a globally flat 
spacetime a fermion field $\psi$ is described using spinor representation of 
the Lorentz group. However in a curved spacetime, such symmetry is available 
only as a local symmetry. In particular, in a curved spacetime, one can always 
find a set of \emph{local} coordinates, denoted as $\xi^a$, such that the global 
metric $g_{\mu\nu}$ can be expressed as $g_{\mu\nu} {e^{\mu}}_a {e^{\nu}}_b = 
\eta_{ab}$ where $\eta_{ab} = diag(-1,1,1,1)$ is the Minkowski metric. Here 
${e^{\mu}}_a$ are the \emph{tetrad} components defined as ${e^{\mu}}_a \equiv 
({\partial x^{\mu}}/{\partial \xi^{a}})$ where we denote indices of the global 
coordinates $x^{\mu}$ using the Greek letters and indices of the locally 
inertial coordinates $\xi^a$ using the Latin letters. By using the tetrad and 
the \emph{inverse} tetrad ${e_{\mu}}^{a}$, we can relate components of a vector 
field in the global frame to components of the corresponding vector field in the 
local frame.

The coordinate transformation $\xi^{a}\rightarrow \xi'^{a}$ generates a local 
Lorentz transformation ${\Lambda^a}_b(x) = (\partial\xi'^a/\partial\xi^b)$ under 
which a fermion field $\psi(x)$ transforms  as $\psi (x)\rightarrow \psi'(x) = 
U[\Lambda(x)]\psi(x)$ where $U[\Lambda] = \mathbb{I} + \frac{1}{8} \Omega_{ab} 
[\gamma^{a},\gamma^{b}]$ with $\Omega_{ab}$ being the transformation parameters. 
Here $\gamma^{a}$ are the Dirac matrices in Minkowski spacetime and satisfy the 
Clifford algebra $\{\gamma^{a}, \gamma^{b}\} = - 2\eta^{ab} \mathbb{I}$. We have 
used minus sign in front of $\eta^{ab}$ such that for its given signature the 
usual relations $(\gamma^0)^2 = \mathbb{I}$ and $(\gamma^k)^2 = -\mathbb{I}$ for 
$k=1,2,3$ holds true. Under a local Lorentz transformation, the term 
$\partial_{a}\psi$ however does not transform as a co-vector  \emph{i.e.} 
$\partial_{a}\psi$ $\rightarrow$ $\partial_{a}'\psi' \ne 
{({\Lambda^{-1})}_{a}}^b U[\Lambda] \partial_{b} \psi$ as 
$\partial_{\mu}U[\Lambda(x)] \ne 0$ in a generally curved spacetime. So in order 
to formulate the Dirac action in a curved spacetime, we need to define a 
suitable covariant derivative for the fermion field $\psi$ as
\begin{equation}\label{CovariantDerivative}
\tilde{\mathcal{D}}_{a} \psi \equiv {e^{\mu}}_{a} \mathcal{D}_{\mu}\psi \equiv 
{e^{\mu}}_{a} [\partial_{\mu}\psi + \Gamma_{\mu}\psi] ~,
\end{equation}
such that it transforms as $\tilde{\mathcal{D}}_{a}\psi \rightarrow 
{({\Lambda^{-1})}_{a}}^b U[\Lambda] ~\tilde{\mathcal{D}}_{b}\psi$ under a local 
Lorentz transformation. The covariant derivative acts on the Dirac 
adjoint $\bar{\psi} = \psi^{\dagger} \gamma^0$ as $\mathcal{D}_{\mu}\bar{\psi} 
\equiv  [\partial_{\mu}\bar{\psi} - \bar{\psi}\Gamma_{\mu}]$. The spin 
connection $\Gamma_{\mu}$ can be expressed as $\Gamma_{\mu} = 
-\tfrac{1}{8}\omega_{\mu ab} [\gamma^{a}, \gamma^{b}]$ and by demanding  
compatibility conditions for the tetrads \emph{i.e.} $\mathcal{D}_{\mu}  
{e^{\nu}}_{a} = 0 = \mathcal{D}_{\mu} {e_{\nu}}^{a}$ one can express
\begin{equation}\label{OmegaMuabDef}
\omega_{\mu ab}  = \eta_{ac} {e_{\nu}}^c \left[ \partial_{\mu} {e^{\nu}}_b 
+ \Gamma^{\nu}_{\mu\sigma} {e^{\sigma}}_b \right] ~,
\end{equation}
where $\Gamma^{\nu}_{\mu\sigma}$ are the Christoffel connections. Therefore, a 
generally invariant action for a minimally coupled Dirac field $\psi$ can be 
written as
\begin{equation}\label{FermionActionI}
S_{\psi} = -\int d^{4}x \sqrt{-g} ~ \bar{\psi} [i \gamma^{a} {e^{\mu}}_a 
\mathcal{D}_{\mu} + m]\psi  ~,
\end{equation}
where $m$ is its mass. The corresponding field equation is $[i \gamma^{a} 
{e^{\mu}}_a \mathcal{D}_{\mu} + m]\psi = 0$. The corresponding conservation 
equation is $\mathcal{D}_{\mu} j^{\mu} = \nabla_{\mu}j^{\mu} = 0$ where 
conserved 4-current density is given by $j^{\mu} = \bar{\psi}\gamma^{a} 
{e^{\mu}}_a \psi$ and $\nabla_{\mu}$ denotes standard covariant derivative in 
the curved spacetime. The Lagrangian density $\mathcal{L}$ corresponding to the 
action (\ref{FermionActionI}) is given by $S_{\psi} = \int d^{4}x \sqrt{-g} 
~\mathcal{L}$.

\section{Spacetime Metric}\label{sec:InteriorSpacetime}

The spacetime metric of a slowly rotating, axially symmetric star can be 
represented, in the \emph{natural units} $c = \hbar =1$, by an invariant line 
element \cite{hartle1967slowly, hartle1968slowly}
\begin{equation}\label{SRMetric}
ds^2 = - e^{2\Phi}dt^2 + e^{2\nu} dr^2 + 
r^2[d\theta^2 +  \sin^2\theta (d\varphi-\omega dt)^2] ~,
\end{equation}
where $\omega$ represents the acquired angular velocity by a freely-falling 
observer from infinity, a phenomena known as the \emph{dragging} of inertial 
frames. If one demands that the metric be regular at the origin and it reduces 
to the flat spacetime at infinity then $\omega$ can only be a function of the 
radial coordinate as $\omega = \omega(r)$ \cite{hartle1967slowly}. The metric 
functions $\Phi=\Phi(r)$, $\nu=\nu(r)$ depend only on the radial coordinate so 
that in the absence of the frame-dragging angular velocity $\omega$, the 
spacetime metric (\ref{SRMetric}) represents a spherically symmetric spacetime. 
The radially varying nature of the function $\Phi(r)$ leads to the phenomena of 
gravitational \emph{time dilation}. The mass and radius of a slowly-rotating 
star can be decomposed into a part that corresponds to a non-rotating 
`spherical' star and a set of perturbative corrections to them which are 
$\mathcal{O}(\omega^2)$. Here we shall include the effect of frame-dragging on 
the matter EOS only up to linear order in $\omega$. Henceforth we shall ignore 
all $\mathcal{O}(\omega^2)$ contributions. The \emph{exterior} vacuum Einstein 
equation corresponding to the metric (\ref{SRMetric}) can be solved exactly as
\begin{equation}\label{VacuumSolution}
e^{2\Phi} =  e^{-2\nu} = 1 - \frac{2 G M}{r} ~;~
\omega = \frac{2 G J}{r^3} ~,
\end{equation}
where constant $J$ and $M$ represent the angular momentum and the 
`spherical' mass of the star.

In order to study the \emph{interior} spacetime, one considers the stellar 
matter to be described by a perfect fluid with the stress-energy tensor
\begin{equation}\label{StressEnergyTensorPerfectFluid}
T_{\mu\nu} = (\rho + P) u_{\mu} u_{\nu} + P g_{\mu\nu} ~,
\end{equation} 
where $u^{\mu}$ is the 4-velocity of the stellar fluid satisfying 
$u_{\mu}u^{\mu}= -1$, $\rho$ is the energy density and $P$ is the pressure 
of the fluid. With respect to an observer at infinity if the stellar fluid is 
rotating with an uniform angular velocity $\frac{d\varphi}{dt} = \Omega$, then 
components of the 4-velocity $u^{\mu}$ are related as $u^{\varphi} = \Omega 
u^{t}$ such that $u^{\mu} = (e^{-\Phi}~,0,0,\Omega e^{-\Phi})$. The 
non-vanishing components of its co-vector $u_{\mu}$ are
\begin{equation}\label{Fluid4Velocity}
u_t = -e^{\Phi},~ u_{\varphi} = r^2\sin^2\theta (\Omega-\omega) e^{-\Phi} ~.
\end{equation}
Then the Einstein equation for the metric (\ref{SRMetric}) leads to the 
following equations for $\nu$
\begin{equation}\label{TOVMEqn}
\ e^{-2\nu} = 1 - \frac{2 G \mr}{r} ~,~
\frac{d\mr}{dr} = 4\pi r^{2}\rho ~.
\end{equation}
The metric function $\Phi$ and the pressure $P$ are governed by the equations
\begin{equation}\label{TOVEqn}
\frac{d\Phi}{dr} = \frac{G(\mr + 4\pi r^3 P)}{r(r - 2 G \mr)} ~~,~~
\frac{dP}{dr} = - (\rho + P) \frac{d\Phi}{dr} ~.
\end{equation}
The equation for the frame-dragging angular velocity $\omega$ follows from the 
$t-\varphi$ component of the Einstein equation and is given by
\begin{equation}\label{OmegaEqn}
\frac{1}{r^{4}}\frac{d}{dr}\left(r^{4}j\frac{d \omega}{dr}\right) + 
\frac{4}{r}\frac{dj}{dr}(\omega - \Omega) = 0 ~,
\end{equation}
where $j=e^{-(\nu+\Phi)}$. The solutions of the differential equations 
(\ref{TOVMEqn}, \ref{TOVEqn}, \ref{OmegaEqn}) are subject to the boundary 
conditions $e^{2\Phi(R)} = (1 - 2 G M/R)$ and $\omega'(R) = -3\omega(R)/R$ where 
$\omega' = (d\omega/dr)$ and $R$ denotes the radius of the `spherical' part. 
Additionally, regularity of the equation (\ref{OmegaEqn}) at $r=0$, demands 
$\omega'(0) = 0$. We may mention that the inequality $\omega < \Omega$ holds 
everywhere and a slowly rotating star here implies $\Omega R \ll 1$.

\section{Equation of State}

The Einstein equation leads to \emph{four} independent equations (\ref{TOVMEqn}, 
\ref{TOVEqn}, \ref{OmegaEqn}) for \emph{five} unknown functions namely $(\mr, 
\Phi, P, \rho, \omega)$. So for solving these equations consistently, an 
additional relation between the pressure $P$ and the energy density $\rho$ needs 
to be provided in the form of an equation of state $P = P(\rho)$. In order to 
compute the EOS in the curved spacetime here we follow the methods of thermal 
quantum field theory \cite{laine2016basics, kapusta1989finite, das1997finite}.

\subsection{Stress-energy tensor from partition function}

The perfect fluid form of the stress-energy tensor 
(\ref{StressEnergyTensorPerfectFluid}) implies that the pressure $P$ and the 
energy density $\rho$ can be expressed as
\begin{equation}\label{PressureEnergyDensityGen}
P = \frac{1}{3} h^{\mu\nu} T_{\mu\nu} ~,~ \ \rho = u^{\mu}u^{\nu} T_{\mu\nu} ~,
\end{equation}
where $h^{\mu\nu} = g^{\mu\nu} + u^{\mu}u^{\nu}$ projects any 4-vectors to the 
hyper-surface orthogonal to $u^{\mu}$. On the other hand, the stress-energy 
tensor corresponding to the Dirac action (\ref{FermionActionI}) can be written 
as
\begin{equation}\label{StressEnergyTensorFull}
T_{\mu\nu} = -
\frac{e_{(\mu a}}{\sqrt{-g}}\frac{\delta S_{\psi}}{\delta e_{ \ a}^{\nu)}}  
~=~ \bar{\psi} [i \gamma^{a} {e_{(\mu a}} \mathcal{D}_{\nu)}] \psi
+ g_{\mu\nu}\mathcal{L}  ~.
\end{equation}
Using the equations (\ref{PressureEnergyDensityGen}, 
\ref{StressEnergyTensorFull}), we can express energy density $\rho$ as
\begin{equation}\label{EnergyDensityPFDef}
\rho = -\mathcal{L} + \bar{\psi} [i \gamma^{a} 
 u^{\mu}u^{\nu} {e_{(\mu a}} \mathcal{D}_{\nu)} ] \psi  ~,
\end{equation}
and the pressure $P$ as
\begin{equation}\label{PressurePFDef}
P = \mathcal{L} + \frac{1}{3} \bar{\psi} [i \gamma^{a} 
 h^{\mu\nu} {e_{(\mu a}} \mathcal{D}_{\nu)} ] \psi  ~.
\end{equation}
In the framework of quantum field theory in curved spacetime, the stress-energy 
tensor (\ref{StressEnergyTensorPerfectFluid}) should be viewed as an expectation 
value of the corresponding quantum operator, as $T_{\mu\nu} = \langle 
\hat{T}_{\mu\nu}\rangle$. In order to compute the expectation value, here we 
follow the methods of thermal quantum field theory as pioneered by Matsubara 
\cite{matsubara1955new}. The partition function that describes a thermal 
system in equilibrium is given by
\begin{equation} \label{PartionFunctionDef}
\mathcal{Z}_{\psi} = \text{Tr}\left[e^{-\beta(\hat{H} - \mu\hat{N})}\right] ~,
\end{equation}
where $\beta = 1/k_{B}T$ with $T$ being the temperature of the system 
and $k_{B}$ is the Boltzmann constant. The number operator of the fermions is
$\hat{N} = \int d^3x \sqrt{-g} ~\hat{n}$ where $\hat{n} = j^t = \bar{\psi} 
\gamma^{a} {e^{t}}_a \psi$ is the number density operator and $\mu$ is the 
associated chemical potential. The Hamiltonian operator is $\hat{H} = \int d^3x 
\sqrt{-g} ~\mathcal{\hat{H}}$  where Hamiltonian density is $\mathcal{\hat{H}} = 
-\bar{\psi}[i \gamma^{a} {e^t}_a \partial_t ]\psi -\mathcal{L}$. Using the 
partition function (\ref{PartionFunctionDef}), it is straightforward to arrive 
at the following expression of the number density
\begin{equation}\label{NumberAndPressureDef}
n =\langle \hat{n}\rangle = \frac{1}{\beta V} 
\frac{\partial\ln\mathcal{Z}_{\psi}}{\partial\mu}  ~,
\end{equation}
where $V = \int d^3x\sqrt{-g}$ is the volume of the system. Then the energy 
density $\rho$ (\ref{EnergyDensityPFDef}) can be expressed as
\begin{equation}\label{EnergyDensityEVDef}
\rho -\mu n + \frac{1}{V}\frac{\partial\ln\mathcal{Z}_{\psi}}{\partial\beta} 
= \langle \bar{\psi} [i \gamma^{a} ( {e^t}_a\partial_t +
 u^{\mu}u^{\nu} {e_{(\mu a}} \mathcal{D}_{\nu)} )]\psi \rangle ~,
\end{equation}
and the pressure $P$ (\ref{PressurePFDef}) can be expressed as 
\begin{equation}\label{PressureEVDef}
3 P =  \rho +  \frac{m}{\beta V} 
\frac{\partial\ln\mathcal{Z}_{\psi}}{\partial m}  
+ 3 \langle\mathcal{L}\rangle ~.
\end{equation}

\subsection{Metric within a box}

Inside a star both the pressure and the energy density vary radially in 
general. On the other hand, at a thermal equilibrium, these quantities are 
uniform within a thermal ensemble. So in order to combine these two aspects 
together, one needs to consider a \emph{small} enough spatial region around each 
point within the star such that variation of the metric within the region can be 
ignored yet it contains sufficiently \emph{large} number of degrees of freedom. 
In other words, for a consistent description of the pressure and the energy 
density, the notion of \emph{local} thermodynamical equilibrium must hold inside 
a star. Therefore, in quantum statistical physics the many-particle 
wave-function must also be localized within the small region to ensure local 
thermodynamical equilibrium. Besides, a large number of degrees of freedom  
makes the fluctuations of the observables much smaller compared to their 
averaged values.

Let us now consider a small box located at the coordinate $(r_{0},\theta_0)$ 
inside the star. In order to ensure local thermal equilibrium, we consider 
the metric inside the box to be uniform. By defining a new set of coordinates $X 
= e^{\nu(r_0)} r \sin \bar{\theta} \cos\bar{\varphi}$, $Y = e^{\nu(r_0)} r \sin 
\bar{\theta} \sin\bar{\varphi}$, and $Z = e^{\nu(r_0)} r \cos \bar{\theta}$ 
along with $\bar{\theta} = e^{-\nu(r_0)}\theta$, $\bar{\varphi} = \xi_0 \varphi$ 
and $\xi_0 = e^{-\nu(r_0)}\sin\theta_0/\sin(e^{-\nu(r_0)}\theta_0)$, the metric 
within the box becomes
\begin{equation}\label{MetricInBox}
g_{\mu\nu} = \begin{bmatrix}
-e^{2\Phi} & \omega Y & - \omega X & 0\\
\omega Y & 1 & 0 & 0\\
- \omega X & 0 & 1 & 0\\
0 & 0 & 0 & 1
\end{bmatrix}  ~,
\end{equation}
where $\Phi = \Phi(r_0)$, $\omega = \omega(r_0)$. We have ignored 
$\mathcal{O}(\omega^2)$ terms here as the star is slowly rotating. To arrive at 
the metric (\ref{MetricInBox}), here we have approximated $\sin^2\theta d\varphi 
= \sin^2(\theta_0 +\delta\theta) d\varphi \approx \sin^2\theta_0 d\varphi$ for 
all points inside the small box. Additionally, we have expanded $\xi_0 = 1 + 
\Delta(r_0, \theta_0)$ and kept only the leading term  as the sub-leading terms 
are expected to contribute perturbatively to the `non-spherical' part of the 
slowly rotating star. The metric within the box (\ref{MetricInBox}) retains the 
information about the metric functions $\Phi$ and $\omega$, in contrast to the 
usage of a globally flat spacetime for computing the matter EOS in the 
literature \cite{shen2002complete, douchin2001unified, lattimer2016equation, 
tolos2016equation, ozel2016dense, katayama2012equation}. These metric functions 
are treated as constants within the \emph{scale of the box}, a scale which is 
sufficient to describe  the microscopic physics. However, these metric 
functions vary at the \emph{scale of the star}, as governed by the equations 
(\ref{TOVEqn}, \ref{OmegaEqn}).

\subsection{Reduced fermion action}

In this section, we derive a reduced action that describes an ensemble of  
non-interacting Dirac fermions contained in the given box. Corresponding to the 
metric (\ref{MetricInBox}), non-vanishing components of the Christoffel 
connection are given by
\begin{equation}\label{ChristoffelInBox}
\Gamma^X_{tY} = \Gamma^X_{Yt} = \omega  ~,~
\Gamma^Y_{tX} = \Gamma^Y_{Xt} = -\omega  ~.
\end{equation}
Similarly for the metric (\ref{MetricInBox}), the non-vanishing tetrad 
components are ${e^X}_1 = {e^Y}_2 = {e^Z}_3 = 1$, and 
\begin{equation}\label{TetradInBox}
{e^t}_0 = e^{-\Phi}  ~,~
{e^X}_0 = -\omega Y e^{-\Phi} ~,~
{e^Y}_0 =  \omega X e^{-\Phi}  ~.
\end{equation}
On the other hand, non-vanishing components of the inverse tetrad can be 
found as ${e_X}^1 = {e_Y}^2 = {e_Z}^3 = 1$ and 
\begin{equation}\label{InverseTetradInBox}
{e_t}^0 = e^{\Phi}  ~,~
{e_t}^1 = \omega Y  ~,~
{e_t}^2 = -\omega X  ~.
\end{equation}
Consequently, the following components of $\omega_{\mu ab}$ (\ref{OmegaMuabDef}) 
are non-vanishing 
\begin{equation}\label{OmegaMuabInBox}
\omega_{t12} = - ~\omega_{t21} = \omega   ~.
\end{equation}
From the equations (\ref{OmegaMuabDef}, \ref{OmegaMuabInBox}), we note that the 
only non-vanishing component of the spin-connection $\Gamma_{\mu}$ is
\begin{equation}\label{SpinConnectionInBox}
 \Gamma_{t} = - \frac{\omega}{4}[\gamma^{1},\gamma^{2}] = 
\frac{i \omega}{2}\sigma^3 \otimes\mathbb{I}_2  ~.
\end{equation}
In the equation (\ref{SpinConnectionInBox}), we have used following 
representation of the Dirac matrices
\begin{equation}\label{DiracMatrices}
\gamma^{0} = \begin{bmatrix}
\mathbb{I}_{2} & 0\\
0 & - \mathbb{I}_{2}
\end{bmatrix}, \ \gamma^{k} = \begin{bmatrix}
0 & \sigma^{k}\\
-\sigma^{k} & 0
\end{bmatrix},
\end{equation}
where $\sigma^{k}$ with $k=1,2,3$ are the Pauli matrices. Consequently, within 
the box the Dirac action (\ref{FermionActionI}) reduces to
\begin{equation}\label{FermionActionInBox}
S_{\psi} = -\int d^{4}x ~\bar{\psi} [ i \gamma^0 \partial_t + 
e^{\Phi} (i\gamma^k \partial_k + m) - \omega \gamma^0 \hat{J}_Z ]\psi  ~,
\end{equation}
where $\hat{J}_Z = (\hat{L}_Z + \tfrac{1}{2} \Sigma_3)$ with $\hat{L}_Z = 
-i(X\partial_{Y} - Y\partial_{X})$ and $\Sigma_3 = \sigma^3 \otimes 
\mathbb{I}_2$. We note that $\hat{J}_Z$ can be naturally interpreted as the 
total angular momentum operator arising due to the frame-dragging angular 
velocity $\omega$ where $\hat{L}_Z$ is the orbital angular momentum operator and 
$\sigma^3$ is the third Pauli matrix which is the spin operator along 
$Z$-direction.

Further we note that if the frame-dragging angular velocity $\omega$ were zero 
then the reduced action (\ref{FermionActionInBox}) would become the same as the 
one studied for a spherical star \cite{hossain2021equation, hossain2021higher}. 
Now in a such spherically symmetric spacetime, if one applies a rotation to 
the given box with an angular velocity $\omega$ around $Z$ axis then the 
fermion field would transform as $\psi(x) \mapsto e^{i\hat{J}_{Z}\omega 
t}\psi(x)$. One may check that such a procedure would also led to the same 
reduced action as in (\ref{FermionActionInBox}).

\subsection{Evaluation of partition function}

In the functional integral formulation, using the coherent states of the 
Grassmann fields \cite{laine2016basics, kapusta1989finite, das1997finite}, 
the partition function is expressed as $\mathcal{Z}_{\psi} = \int\mathcal{D}
\bar{\psi} \mathcal{D}\psi \ e^{-S_{\psi}^{\beta}}$ where $S_{\psi}^{\beta} 
= \int_{0}^{\beta}d\tau\int d^{3}x (\mathcal{L}^{E} - \mu\bar{\psi}\gamma^{0}
\psi)$ with $\mathcal{L}^{E}$ being the Euclidean Lagrangian density and is 
obtained through a Wick rotation $\mathcal{L}^{E} = -\mathcal{L}(t\rightarrow 
-i\tau)$. Here $T = 1/(k_B \beta)$ denotes the temperature of the ensemble of 
fermions in the box which is in a local thermodynamical equilibrium. We define 
the scale of temperature $T$ with respect to an  asymptotic observer in whose 
frame $\Phi\to 0$ and $\omega\to 0$. It allows us to treat the reduced 
action (\ref{FermionActionInBox}) as an \emph{effective} action written in the 
Minkowski spacetime. Consequently it leads to a simpler computation of the 
partition function that we shall follow now onward. Furthermore, it allows one 
to avoid the issues related to the Wick rotation  \cite{visser2017wick} or 
metric density dependence of the path integral measure 
\cite{toms1987functional} that would arise in an arbitrary curved spacetime.

In order to evaluate the partition function, it is convenient to split it as
$\ln\mathcal{Z}_{\psi} = \ln\mathcal{Z}_0 + \ln\mathcal{Z}_L$ where 
$\mathcal{Z}_0 = \int\mathcal{D}\bar{\psi} \mathcal{D}\psi ~e^{-S^{\beta}_0}$ 
with 
\begin{equation}\label{EuAction0}
S^{\beta}_0 = \int_{0}^{\beta}d\tau\int d^{3}x \bar{\psi}\big[
-\gamma^{0}(\partial_{\tau} + \mu + \frac{\omega}{2}\Sigma_3) 
+ e^{\Phi} (i\gamma^{k}\partial_{k} + m)\big]\psi .
\end{equation}
On the other hand, $\ln\mathcal{Z}_L$ contains contributions from the orbital 
angular momentum operator $\hat{L}_Z$ and can be expressed as a perturbative 
series
\begin{equation}\label{LnZL}
\ln\mathcal{Z}_L = \ln \left( 1 + \sum_{l=1}^{\infty} \frac{\omega^l}{l!}
\langle (-S^{\beta}_L)^l \rangle \right) ~,
\end{equation}
where $S_{L}^{\beta} =  \int_{0}^{\beta}d\tau\int d^{3}x \bar{\psi}[\gamma^{0}
\hat{L}_{Z}]\psi$. The information about the equilibrium temperature of the 
system is carried through the \emph{anti-periodic} boundary condition of the 
fermion field $\psi$ as
\begin{equation}\label{FermionicBoundaryCondition}
\psi(\tau,\x) = -\psi(\tau+\beta,\x)  ~.
\end{equation}
By using the Matsubara frequencies $\omega_l = (2l+1) \pi/\beta$ where
$l$ is an integer, we can express the field $\psi$ in Fourier domain as
\begin{equation}\label{FermionicFourier}
\psi(\tau,\x) = \frac{1}{\sqrt{V}} \sum_{l,\k} ~e^{-i(\omega_l\tau + 
\k\cdot\x)} \tilde{\psi}(l,\k)  ~,
\end{equation}
where volume of the box is now $V = \int d^3x \sqrt{-\eta}$. The equation 
(\ref{FermionicFourier}) then leads the action (\ref{EuAction0}) to become
\begin{equation}\label{EuAction0II}
S^{\beta}_0 = \sum_{l,\k} ~\bar{\tilde{\psi}}~\beta
\left[ \slashed{p} + \bar{m} \right]  \tilde{\psi} ~,
\end{equation}
where $\bar{m} =  m e^{\Phi}$,  $\slashed{p} = 
\gamma^{0}(i\omega_l - \mu - \frac{\omega}{2} \Sigma_3) + 
\gamma^{k} (\k_k e^{\Phi})$. 
From the equation (\ref{EuAction0II}), we can read off the corresponding 
thermal propagator in Fourier domain as
\begin{equation}\label{SpinorPropagator}
\mathcal{G}(\omega_l,\k) = \frac{1}{\slashed{p} + \bar{m}} ~.
\end{equation}
The term $\ln\mathcal{Z}_L$ can be computed either by using the equation
(\ref{FermionicFourier}) or in principle  by following the approach in 
\cite{ambrucs2014rotating, ambrucs2016rotating, chernodub2017effects, 
iyer1982dirac, vilenkin1980quantum}. In any case, it can be shown that the 
leading order terms in $\ln\mathcal{Z}_L$ are $\mathcal{O}(\omega^2)$ which we 
neglect henceforth for a slowly-rotating star.
Using the results of Gaussian integral over Grassmann fields and the Dirac 
representation of $\gamma^a$ matrices, one can evaluate the total partition 
function as
\begin{equation}\label{LogZSplitPM}
\ln\mathcal{Z}_{\psi} = \ln\mathcal{Z}_{-} + \ln\mathcal{Z}_{+} ~,
\end{equation}
where 
\begin{equation}\label{LogPartitionFunctionPM}
\ln\mathcal{Z}_{\pm} = \sum_{l,\k} ~ 
\ln \left[\beta^2 \{(\omega_l + i\mu_{\pm})^2 + \varepsilon^2 \} \right] ~,
\end{equation}
with $\varepsilon^2 = \varepsilon(\k)^2 = (\k^2 + m^2)e^{\Phi}$ and $\mu_{\pm} = 
\mu \pm (\omega/2$). Here we have ignored terms which are 
$\mathcal{O}(\omega^2)$. The presence of Pauli matrix $\sigma^3$ and 
non-vanishing frame-dragging angular velocity $\omega$ in the expression 
(\ref{EuAction0II}) leads to the splitting of partition function 
(\ref{LogZSplitPM}) into two parts corresponding to the contributions from the 
spin-up and the spin-down fermions respectively. This breaking of 
spin-degeneracy of fermions leads to a novel mechanism for generation of seed 
magnetism in spinning astrophysical bodies \cite{SeedMagnetism2022}.

By using the relation $\beta^2 \{(\omega_l + i\mu_{\pm})^2 + \varepsilon^2\} 
= \{(\beta\varepsilon - \beta\mu_{\pm} + i\pi) + i2l\pi\} \{(\beta\varepsilon + 
\beta\mu_{\pm} - i\pi) - i2l\pi\}$  and the identity
\begin{equation}
\frac{\sinh z}{z} = \prod_{l=1}^{\infty} \left(1+\frac{z^2}{l^2\pi^2} \right) ~,
\end{equation}
the summation over $l$ can be carried out which leads to the following 
expression
\begin{equation}\label{LogPartitionFunctionPM2}
\ln\mathcal{Z}_{\pm} = \sum_{\k} \left[ 
\ln\big(1 + e^{-\beta(\varepsilon - \mu_{\pm})} \big) 
+ \ln\big(1 + e^{-\beta(\varepsilon + \mu_{\pm})} \big)
\right] ~.
\end{equation}
To arrive at the expression (\ref{LogPartitionFunctionPM2}), we have dropped 
formally divergent terms including the zero-point energy of fermions. In the 
equation (\ref{LogPartitionFunctionPM2}), the first and the second terms 
correspond to the \emph{particle} and the \emph{anti-particle} sectors 
respectively.

Inside a compact star, the degenerate nature of the fermions can be expressed
by the conditions $\beta\mu_{\pm} \gg 1$. So by using the degeneracy condition 
and by converting the sum into a momentum integral as $\sum_{\k} \to V \int 
\frac{d^3\k}{(2\pi)^3}$ , the expression (\ref{LogPartitionFunctionPM2}) becomes
\begin{equation}\label{LogZPMFinal}
\ln\mathcal{Z}_{\pm}  = \frac{\beta V e^{-3\Phi}}{48\pi^{2}} 
\Big[2\mu_{\pm}\mu_{\pm m}^{3} - 3\bar{m}^{2}\bar{\mu}_{\pm m}^{2}
+ \frac{48\mu_{\pm}\mu_{\pm m}} {\beta^{2}} \Big]  ~, 
\end{equation}
where $\mu_{\pm m} = \sqrt{\mu_{\pm m}^{2} - \bar{m}^{2}}$ and 
$\bar{\mu}_{\pm m}^{2} = \mu_{\pm}\mu_{\pm m} - \bar{m}^{2}
\arcsinh(\mu_{\pm m}/\bar{m})$. 
We note that if one turns off the gravitational time dilation and the dragging 
of inertial frames by setting $\Phi\to 0$ and $\omega\to 0$ respectively, then 
the partition function (\ref{LogZSplitPM})  reduces to the one evaluated in the 
Minkowski spacetime containing standard spin-degeneracy factor of $2$. 
Secondly, if one turns off only the frame-dragging effect by setting $\omega\to 
0$, then the partition function (\ref{LogZSplitPM}) reduces to the one computed 
in the curved spacetime of a spherical star \cite{hossain2021equation}. It 
follows from the fact that $\ln\mathcal{Z}_{+} = \ln\mathcal{Z}_{-}$ in the 
absence of the frame-dragging angular velocity $\omega$.

\subsection{Pressure and energy density}

Using the equation (\ref{NumberAndPressureDef}) together with the partition 
function (\ref{LogZSplitPM}) and ignoring temperature dependent small 
corrections, we obtain following expression for the number density $n$ as
\begin{equation}\label{NumberDensityPM}
n = n_{+} + n_{-} ~,~~  \text{with} ~~
n_{\pm} = \frac{e^{-3\Phi}}{6\pi^{2}} \mu_{\pm m}^{3}  ~.
\end{equation}
We may mention here that $\mu_{\pm}$ can be equivalently treated as independent 
variables in places of $\mu$ and $\omega$. The equation (\ref{NumberDensityPM}) 
can be used to express chemical potentials in terms of the respective number 
densities as
\begin{equation}\label{MuUsingNumberDensityPM}
\mu_{\pm} = m e^{\Phi} \sqrt{(bn_{\pm})^{2/3} + 1} ~,~~
\text{with} ~~ \mu_{+} - \mu_{-} = \omega ~,
\end{equation}
where the constant $b = (6\pi^2/m^3)$. For a non-vanishing $\omega$, the 
equation (\ref{MuUsingNumberDensityPM}) cannot be satisfied below a threshold 
number density $n_0 = b^{-1}(2\omega e^{-\Phi}/m)^{3/2}$ where $n_{-}$ becomes 
zero. It turns out that $(\omega/m)$ is an extremely small number for any 
regular degenerate star. Therefore, the assumed degeneracy condition is
expected to fail much above the threshold number density $n_0$. Additionally, 
near the surface of the star where the number density becomes very low, it is 
desirable to use a different matter EOS, rather than the one for a degenerate 
matter.

With the metric (\ref{MetricInBox}), the 4-velocity of the stellar fluid in the 
box can be obtained as  $u^{\mu} = e^{-\Phi}(1, -\omega Y, \omega X, 0)$ along 
with  its co-vector $u_{\mu} = e^{\Phi}(-1, 0, 0, 0)$. It leads to the projector 
$h^{\mu\nu} = diag(0, 1, 1, 1)$. Consequently, the energy density and the 
pressure in the box can be expressed as
\begin{equation}\label{EnergyDensityPressureInBox}
\rho -\mu n + \frac{1}{V}\frac{\partial\ln\mathcal{Z}_{\psi}}{\partial\beta} 
= \frac{\omega}{\beta V}\frac{\partial\ln\mathcal{Z}_{\psi}}{\partial\omega} 
~,~ P = \frac{\ln\mathcal{Z}_{\psi}}{\beta V}  ~.
\end{equation}
In order to arrive at the expression of $P$, we have used on-shell condition
$[i \gamma^{a} {e^{\mu}}_a \mathcal{D}_{\mu} + m]\psi = 0$. Additionally, we 
have used the fact the logarithm of the partition function (\ref{LogZSplitPM}), 
being a dimensionless, \emph{extensive} quantity, can be expressed as 
$\ln\mathcal{Z}_{\psi} = \beta^{-3}V f(\beta\mu, \beta m, \beta\omega)$. 
Following the equation (\ref{EnergyDensityPressureInBox}), we can express total 
pressure as $P = P_{+} + P_{-}$ where
\begin{eqnarray}\label{PressurePM}
P_{\pm} = e^{\Phi} \frac{m^4}{48\pi^2} && \left[ \sqrt{(b n_{\pm})^{2/3} + 1}
\left\{2(b n_{\pm}) - 3(b n_{\pm})^{1/3} \right\} \nonumber \right. \\
&& + \left. ~3 \arcsinh\left\{ (b n_{\pm})^{1/3} \right\} \right]  ~. 
\end{eqnarray}
Similarly, we can express total energy density as $\rho = \rho_{+} + \rho_{-}$ 
where
\begin{equation}\label{EnergyDensityPM}
\rho_{\pm} = - P_{\pm} + e^{\Phi}\frac{m^4}{6\pi^2} 
(b n_{\pm})\sqrt{(b n_{\pm})^{2/3} + 1} ~.
\end{equation}
The equation of state (\ref{PressurePM}, \ref{EnergyDensityPM}) computed in the 
curved spacetime of a slowly rotating star \emph{i.e.} the curved EOS, depends 
explicitly on the metric functions $\Phi$ and $\omega$. Therefore, in contrast 
to the flat EOS, the curved EOS (\ref{PressurePM}, \ref{EnergyDensityPM}) 
captures the effects of both gravitational time dilation and the dragging of 
inertial frames on the matter field dynamics inside a slowly rotating star. As 
expected, the curved EOS reduces to its Minkowski spacetime counterpart in the 
limit $\Phi\to 0$ and $\omega\to 0$. The curved EOS has been computed here for 
a given box located at the coordinates $(r_0,\theta_0)$. However, these 
coordinates being arbitrary the computed curved EOS can be treated as dependent 
on the coordinates through the metric functions $\Phi$ and $\omega$ whose 
dynamics, at the scale of the star, are governed by the equations 
(\ref{TOVMEqn}, \ref{TOVEqn}, \ref{OmegaEqn}).

\section{Numerical Evaluation}

\begin{figure}
\includegraphics[height = 7cm, width = 9cm]{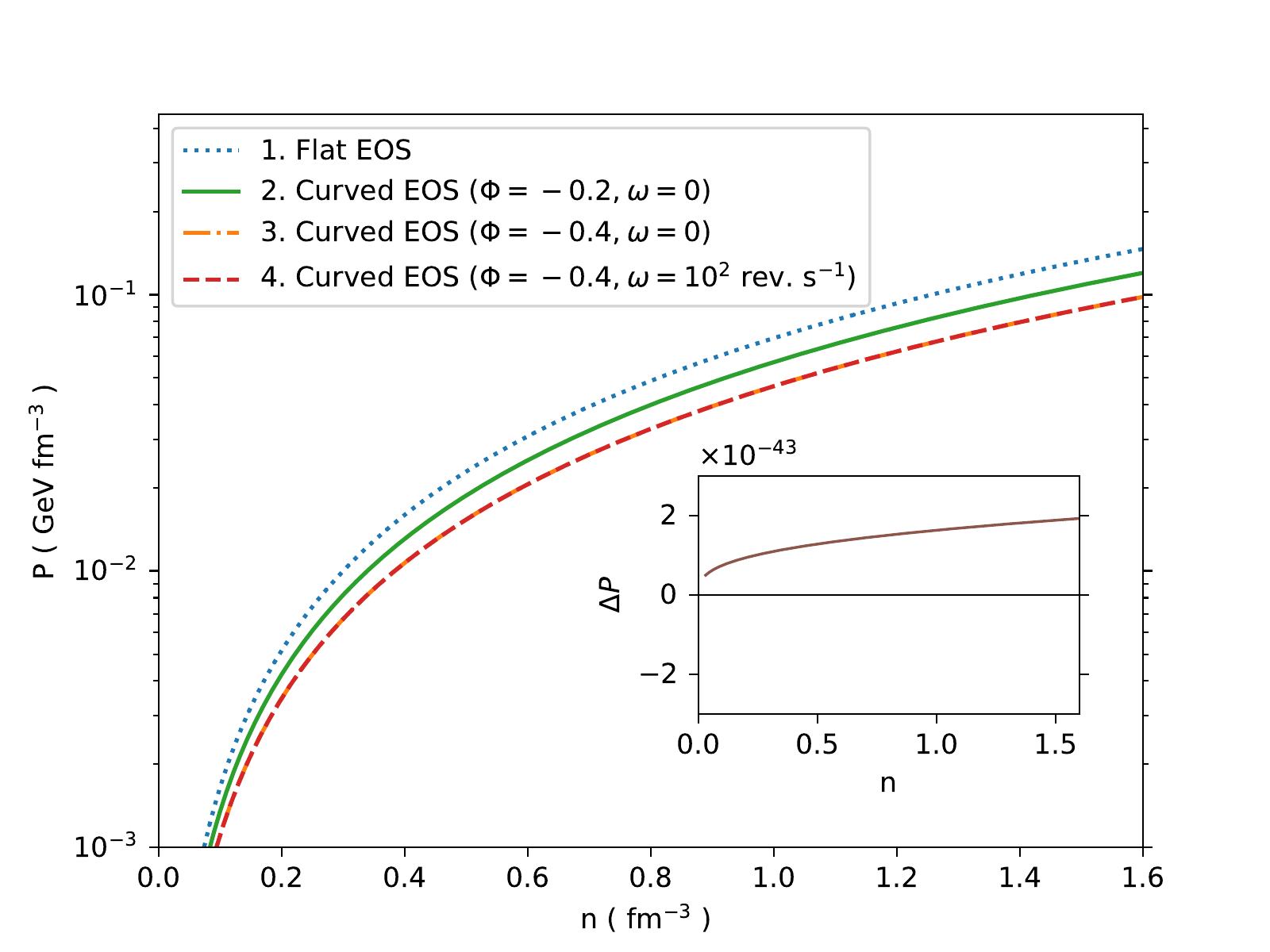}
\caption{Plot of the pressure $P$ exerted by an ensemble of degenerate neutrons 
inside a slowly rotating ideal neutron star as a function of total neutron 
number density $n$ for different kinematical values of $\Phi$ and $\omega$. Here 
$\omega$ is specified in the unit of revolution per second. We note that 
the curves $3$ and $4$ with different values of $\omega$ are effectively 
indistinguishable. The difference between the curves $3$ and $4$ \emph{i.e.} 
$\Delta P \equiv P_4 - P_3$ is shown in the inset plot using same units as the 
main plot. It shows that effect of dragging of inertial frames on the pressure, 
described by $\omega$, is extremely small. In contrast, the effect of 
gravitational time-dilation on the pressure, described by $\Phi$, is 
considerably large.}
\label{fig:pressure-comparison}
\end{figure}

In order to study  implications of the curved EOS (\ref{PressurePM}, 
\ref{EnergyDensityPM}), now we consider an \emph{ideal} neutron star whose 
matter contents are made of non-interacting degenerate neutrons. The EOS for 
such an ensemble of \emph{non-interacting} neutrons was studied first by 
Oppenheimer. Henceforth we consider the parameter $m$ to be the mass of a 
neutron. For such an ideal neutron star, the pressure $P$ is plotted as a 
function of the neutron number density $n$ in the FIG. 
\ref{fig:pressure-comparison}. We may note that the gravitational time dilation 
decreases pressure for a given neutron number density $n$ as $e^\Phi < 1$ inside 
any star. For a given set of values of $n$ and $\Phi$, dragging of inertial 
frames leads to an increase of pressure although by an extremely small amount as 
can be seen in the inset plot of the FIG. \ref{fig:pressure-comparison}.

\begin{figure}
\includegraphics[height = 7cm, width = 9cm]{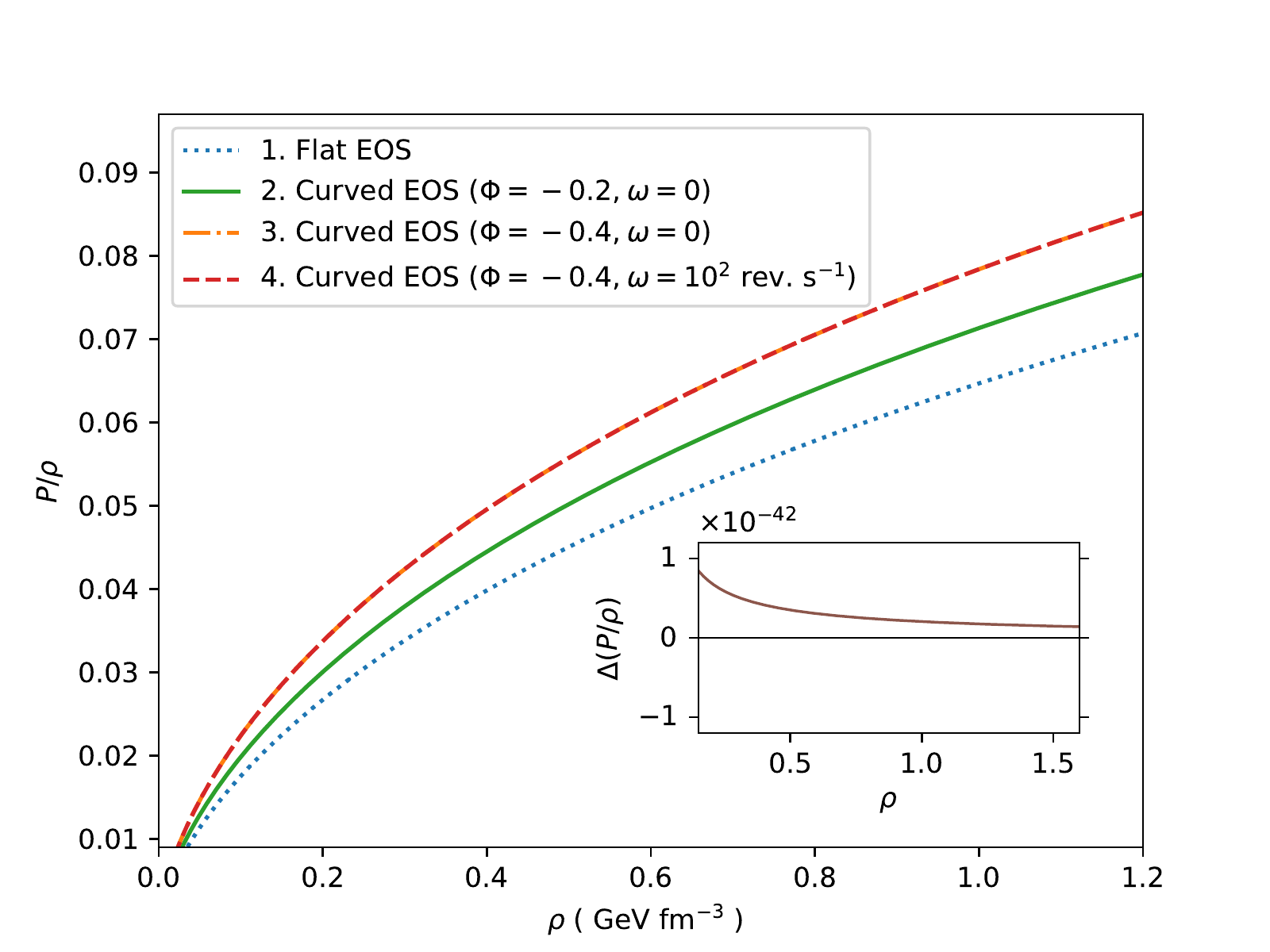}
\caption{Plot of the ratio $P/\rho$ as a function of the energy density $\rho$ 
for different kinematical values of the metric functions $\Phi$ and $\omega$. 
The curves $3$ and $4$ with different values of $\omega$ are indistinguishable 
as earlier. The difference between these two curves \emph{i.e.} $\Delta 
(P/\rho) \equiv (P/\rho)_{\omega} - (P/\rho)_{\omega=0}$ is shown in the inset 
plot using same unit for $\rho$ as the main plot. As earlier, we note that the 
effect of gravitational time-dilation on the equation of state is large whereas 
the effect of frame-dragging is extremely small. }
\label{fig:eos_ratio}
\end{figure}

\begin{figure}
\includegraphics[height = 7cm, width = 9cm]{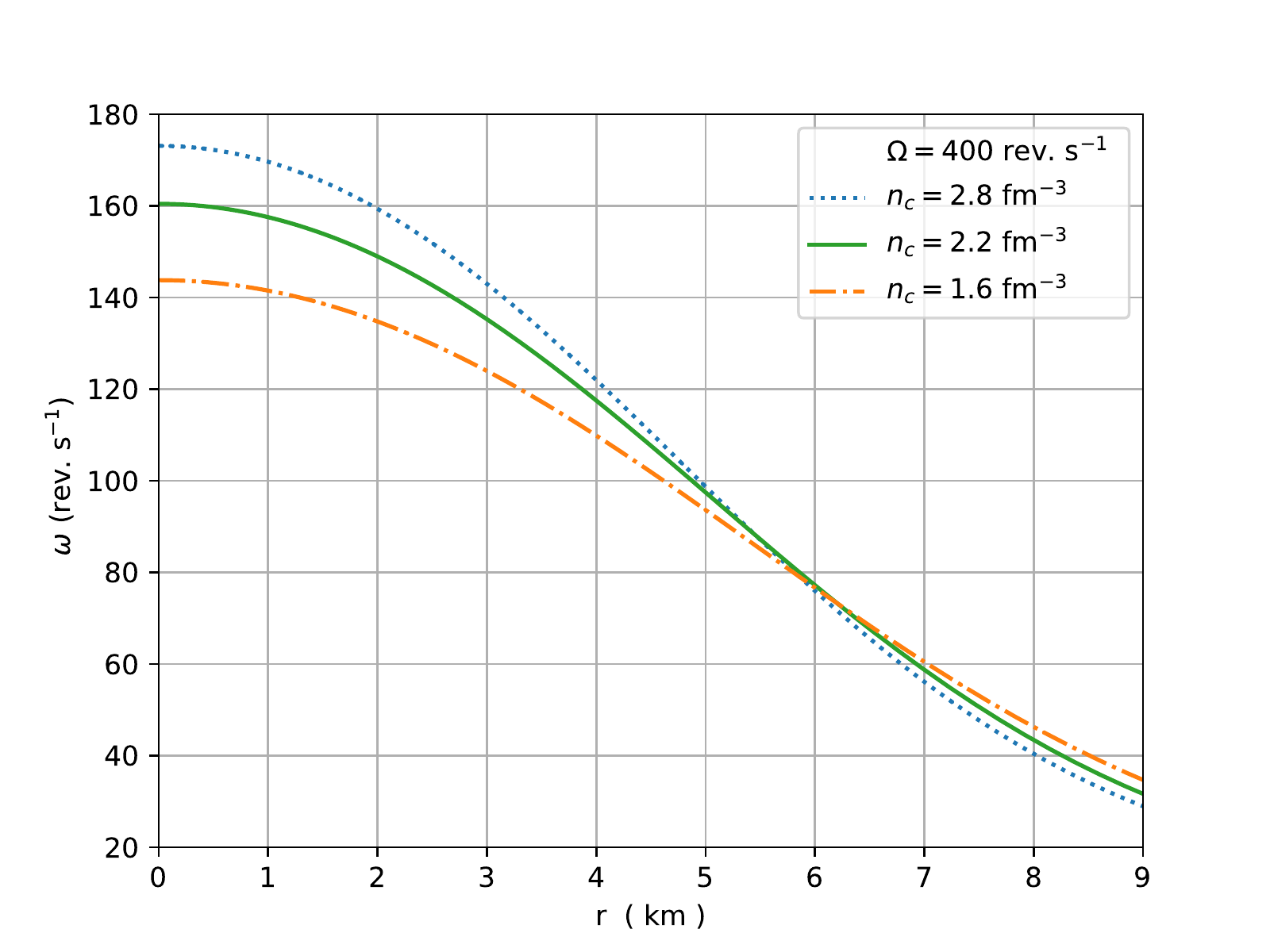}
\caption{The radial variations of frame-dragging angular velocity $\omega$, for 
different values of the central neutron number density $n_c$, inside an ideal 
neutron star whose stellar fluid is revolving $400$ times per second. }
\label{fig:omega-radial}
\end{figure}

However, the Einstein equation is not directly sensitive to the neutron number 
density $n$. Its dependence on the Einstein equation enters through the 
pressure $P$ and the energy density $\rho$. So it is more apt to look at the 
dependence of the pressure $P$ on the energy density $\rho$. The ratio $P/\rho$ 
is plotted as a function of the energy density $\rho$ in the FIG. 
\ref{fig:eos_ratio}. From the figure, we observe that the gravitational time 
dilation as well as the dragging of inertial frames both lead to an enhancement 
of the pressure $P$ for a given energy density $\rho$. In other words, the 
effect of curved spacetime of a slowly rotating star makes a degenerate 
equation of state \emph{stiffer} compared to its counterpart which is computed 
in the Minkowski spacetime. The gravitational time dilation effect, arising due 
to the varying metric function $\Phi$, leads to a significant enhancement of 
the pressure $P$ compared to its flat spacetime counterpart. However, similar 
enhancement due to the dragging of inertial frames, parameterized by the metric 
function $\omega$, is extremely small. We may mention here that the curved EOS 
depends directly on the frame-dragging angular velocity $\omega$ rather than the 
angular velocity $\Omega$ of the stellar fluid. Nevertheless, a non-vanishing 
frame-dragging angular velocity $\omega$ follows from a non-vanishing angular 
velocity of the stellar fluid $\Omega$  through the solution of the Einstein 
equation (\ref{OmegaEqn}) \emph{i.e.} $\omega = \omega(r,\Omega)$. The radial 
variations of $\omega$ for different parameter values are shown in the FIG. 
\ref{fig:omega-radial}.

\subsection{Solution of the Einstein equation}

In general, a rotating star has a shape of an oblate sphere. However, 
a slowly rotating star can be analyzed by decomposing it into a part that 
corresponds to a non-rotating `spherical' star and a set of  perturbative 
corrections to its mass and radius which are $\mathcal{O}(\omega^2)$ 
\cite{hartle1967slowly}. In this article, we have computed the curved EOS which
includes corrections up to $\mathcal{O}(\omega)$. Therefore, we restrict our 
analysis here for the `spherical' part of the star. In other words, in our 
analysis we ignore terms which are $\mathcal{O}(\omega^2)$ or higher.

We note that the curved EOS (\ref{PressurePM}, \ref{EnergyDensityPM}) can be 
viewed as $P = P(n,\Phi,\omega)$ and $\rho = \rho(n,\Phi,\omega)$. However, due 
to the breaking of spin-degeneracy, it is convenient to treat them as $P = 
P(n_{+},\Phi,\omega)$ and $\rho = \rho(n_{+},\Phi,\omega)$. For simplicity, here 
we shall not stitch the curved EOS with a separate EOS for the crust matter 
near the star surface. So in numerical scheme for solving Einstein's equation, 
below the threshold number density $n_0$,  we shall restrict $n_{-}$ to remain 
zero until $n_{+}$ vanishes at the surface thereby defining the radius of the 
`spherical' part of the star $R$, as $n_{+}(R) = 0$. However, for a slowly 
rotating star the threshold number density $n_0$ is extremely small. For 
example, the ratio $(\omega/m) \sim 10^{-22}$ for an ideal neutron star which 
is revolving a hundred times per second. So for a slowly rotating star the 
threshold number density $n_0$ lies way below the numerical precision which is 
used to define $n_{+}(R)$ numerically.

It is convenient to decompose the second order differential equation 
(\ref{OmegaEqn}) into two first order differential equations for $\omega$ and 
$\omega'$. Consequently, the Einstein equation for a slowly rotating star can be 
expressed as a first order differential equation for the tuple 
$\{\mr,\Phi,n_{+},\omega,\omega'\}$ where $n_{+}$ satisfies
\begin{equation}\label{NPEqn}
\frac{dn_{+}}{dr} = - \frac{(\rho + P + ({\partial P}/{\partial \Phi}))}
{({\partial P}/{\partial n_{+}})} \frac{d\Phi}{dr} 
- \frac{({\partial P}/{\partial \omega})}
{({\partial P}/{\partial n_{+}})} \frac{d\omega}{dr}
~.
\end{equation}
In the equation (\ref{NPEqn}), the partial derivatives are evaluated as
\begin{equation}
\left(\frac{\partial P}{\partial\Phi}\right)_{n_{+},\omega} =  P + \omega n_{-}
~,~ \left(\frac{\partial P}{\partial\omega}\right)_{n_{+},\Phi}  = -n_{-} ~,
\end{equation}
together with
\begin{equation}
\left( \frac{\partial P}{\partial n_{+}} \right)_{\Phi,\omega}  = \sum_{s} 
\frac{\partial P_{s}}{\partial n_{s}} \frac{\partial n_{s}}{\partial n_{+}}
~,~ \frac{\partial P_{s} }{\partial n_{s}} =  
\frac{m (b n_{s}) e^{\Phi}}{\mathcal{F}_{s}}  ~,
\end{equation}
where $\mathcal{F}_{s} = 3 (b n_{s})^{1/3} \sqrt{(b n_{s})^{2/3} + 1}$
with $s\in\{+,-\}$, $(\partial n_{+}/\partial n_{+}) = 1$ and
$(\partial n_{-}/\partial n_{+}) = (\mathcal{F}_{-}/\mathcal{F}_{+})$.

In order to evolve the tuple numerically, a set of initial conditions 
$\{\mr=0,\Phi_c,n_{+c},\omega_c,\omega'=0\}$ is chosen at the center of the 
star. While the value for $n_{+c}$ can be chosen independently, the values for 
$\Phi_c$, $\omega_c$ are determined by imposing the relevant constraints. In 
particular, the chosen values for $\Phi_c$ and $\omega_c$ should be such that 
the tuple satisfies the desired boundary conditions as $e^{2\Phi(R)} = (1 - 2 G 
M/R)$, and $\omega'(R) = -3\omega(R)/R$ where $M = \mr(R)$. The values for 
$\Phi_c$ and $\omega_c$ that satisfy these constraints, can be found within the 
desired numerical precision by using suitable bisection methods.

\begin{figure}
\includegraphics[height = 7cm, width = 9cm]{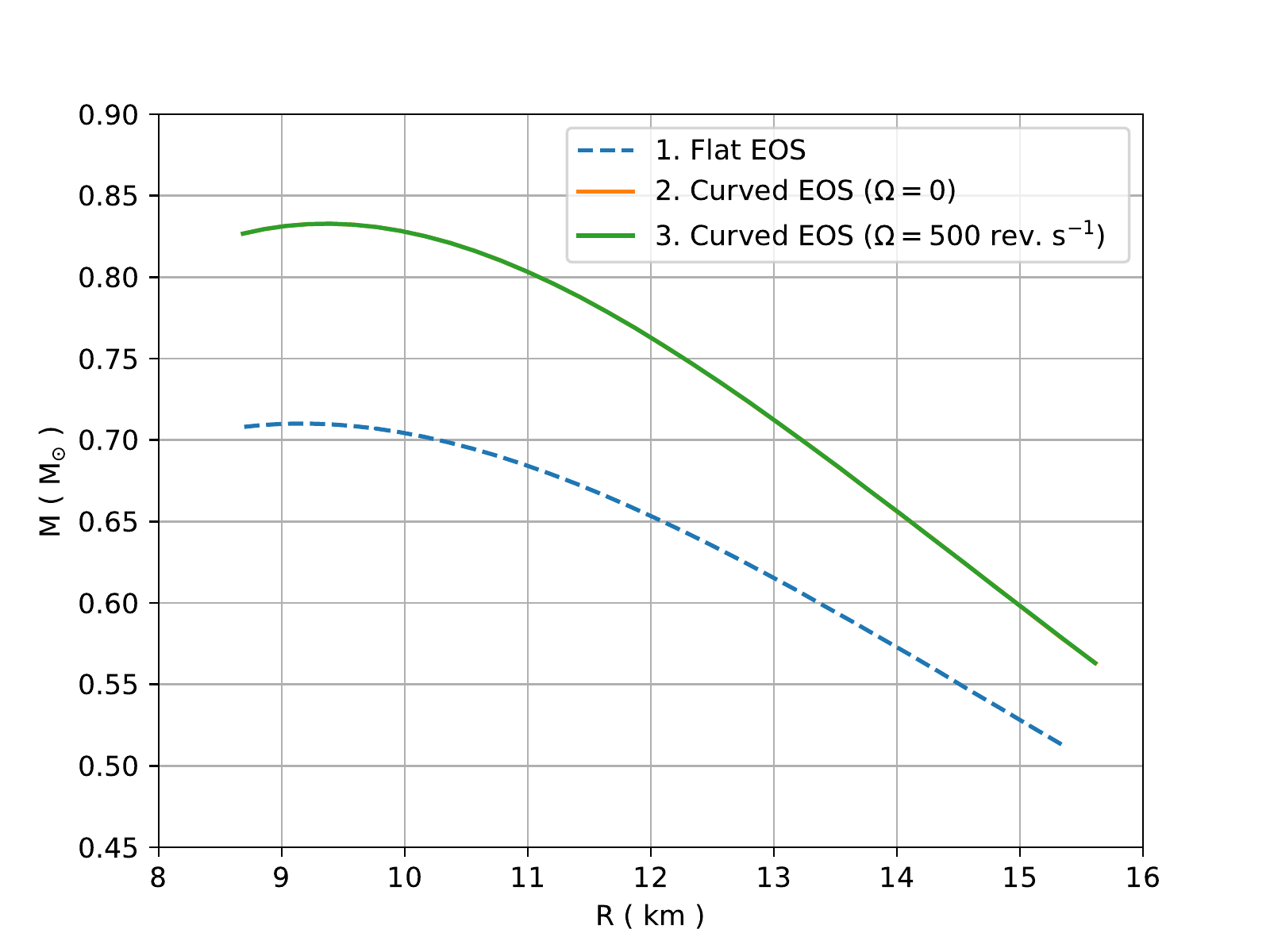}
\caption{Comparison of the mass-radius relations for a slowly-rotating ideal 
neutron star whose degenerate core is made of an ensemble of non-interacting 
neutrons. Usage of the curved EOS leads to a much higher maximum mass limit 
compared to the usage of flat EOS. In particular, the effect of gravitational 
time dilation itself leads to an increase of maximum mass limit from $0.71$ 
M$_{\odot}$ to $0.83$ M$_{\odot}$. The enhancement of mass limit due to the
dragging of inertial frames is however extremely small for any physically 
plausible values of stellar fluid angular velocity $\Omega$. We may note 
that the curves $2$ and $3$ with different values of $\Omega$ are nearly 
indistinguishable. }
\label{fig:mass-radius}
\end{figure}

In the FIG. \ref{fig:mass-radius}, the mass-radius relations for a slowly 
rotating ideal neutron star are plotted, by comparing the results due to the 
flat EOS and the curved EOS. From the figure, it is clear that the effect of 
gravitational time dilation leads to a significant enhancement of the maximum 
mass limit for a slowly rotating neutron star. The dragging of inertial frames 
also leads to an enhancement of the mass limit although by an extremely small 
amount in comparison. The later enhancement nevertheless depends on the angular 
velocity of the stellar fluid $\Omega$. In contrast, the mass-radius relations 
that follow from the flat EOS, has no dependence on the stellar fluid angular 
velocity $\Omega$.

\begin{figure}
\includegraphics[height = 7cm, width = 9cm]{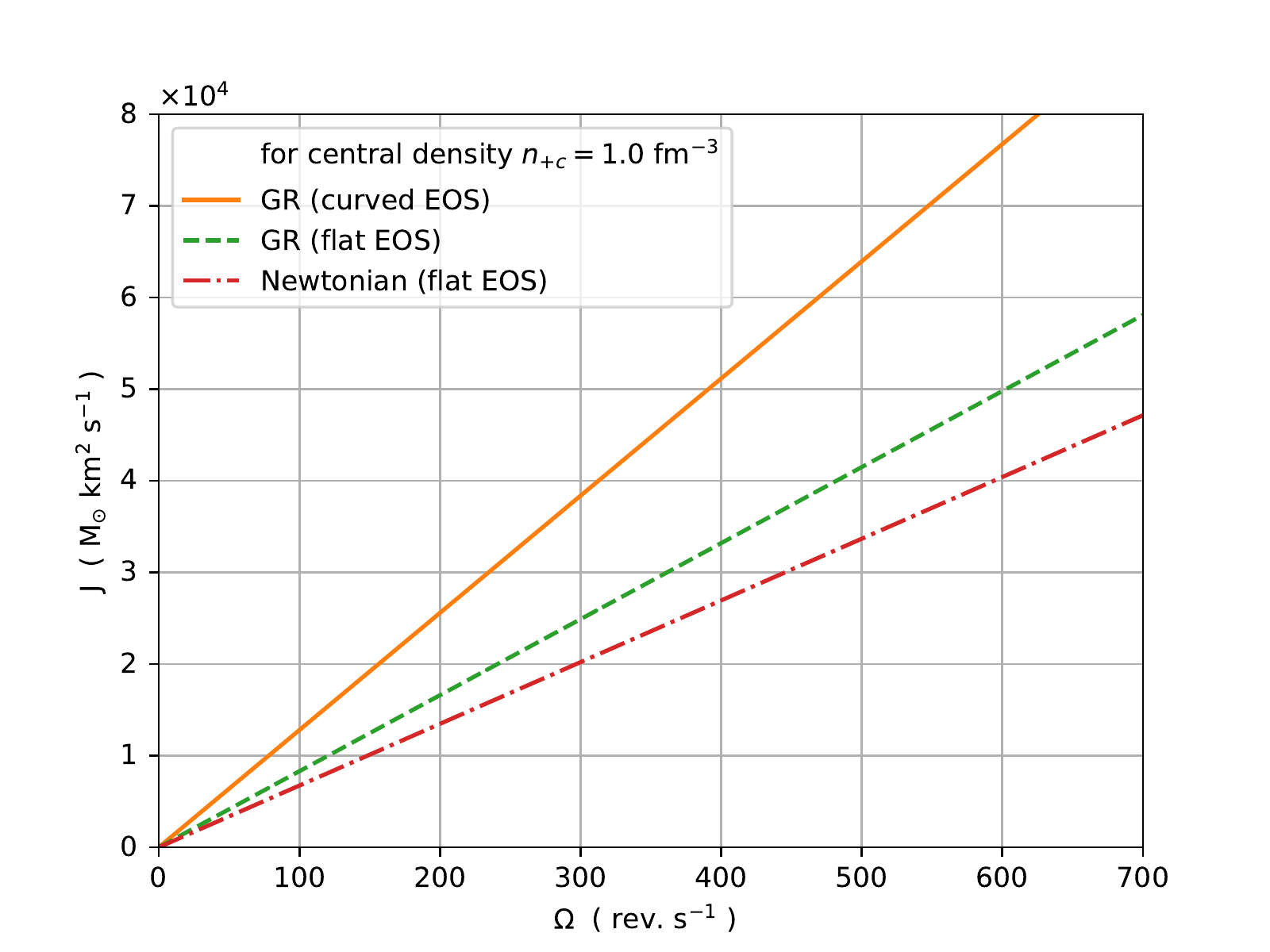}
\caption{Comparison of the angular momenta of a slowly rotating ideal neutron 
star as a function of the angular velocity $\Omega$ of the stellar fluid. The 
Newtonian expression for the angular momenta is computed as $J = \Omega \int 
dr d\theta d\varphi ~ \rho r^4 \sin^3\theta $ where energy density $\rho = 
\rho(r)$ is taken to be the same as found by solving TOV equations with the flat 
EOS. It can be seen that the usage of the curved EOS, rather than the flat EOS, 
leads to higher values of angular momenta for a given value of central number 
density $n_{+c}$.}
\label{fig:angular-momentum}
\end{figure}

From the numerical solution of the tuple, the angular momentum of the star can 
be obtained as $J=\omega(R) R^3/2G$. In the FIG. \ref{fig:angular-momentum}, 
the angular momenta of a slowly rotating ideal neutron star are plotted, by
comparing the usage of the curved EOS and the flat EOS. For a given set of 
values of the central number density $n_{+c}$ and the angular velocity of 
stellar fluid $\Omega$, the usage of curved EOS leads to a higher value for the 
angular momentum.

\subsection{Effects of curved EOS on mass limits}

The maximum mass limits of the compact stars such as the white dwarf 
stars or the neutron stars depend on the detailed nature of the EOS of 
constituent degenerate matter. The well-known Chandrasekhar mass limit of 
$1.4~M_{\odot}$ for the white dwarfs follows from the EOS which is computed by 
considering an ensemble of \emph{non-interacting} degenerate electrons in a flat 
spacetime. By considering a similar approach, Oppenheimer had first shown that 
the maximum mass limit of neutron stars is around $0.71~M_{\odot}$ when one 
uses the EOS for an ensemble of \emph{non-interacting} degenerate neutrons 
computed in a flat spacetime. Here we refer such a neutron star as an 
\emph{ideal} neutron star. However, the mass limit derived by Oppenheimer fails 
to explain the astrophysical neutron stars whose masses are observed to be in 
the range from $1.1~M_{\odot}$ to more than $2~ M_{\odot}$.

In the theoretical approach of Oppenheimer the neutrons within a neutron star 
are taken to be non-interacting but in reality such neutrons are believed to be 
strongly interacting. Unfortunately, the exact nature of nuclear matter 
interaction inside a neutron star is not known and it continues to be an open 
problem due to the incomplete understanding of Quantum Chromodynamics (QCD). So 
in order to produce higher mass limits of neutron stars, different models of 
nuclear matter interactions are considered in the literature.

The aim of the present article however is to study the effects of curved 
spacetime of a slowly rotating star on its degenerate matter EOS, the resultant 
masses and the angular momentum of these stars. In the neutron star literature, 
one usually computes the matter EOS by considering a globally flat spacetime 
rather than using the curved spacetime of the star. As shown here, the usage of 
the curved spacetime of a slowly rotating star contributes two different effects 
on the matter EOS, namely the effects due to the gravitational time-dilation and 
the frame-dragging.

In order to explicitly compute these effects on the matter EOS, for simplicity, 
here we have considered an ideal neutron star whose degenerate core 
consists of non-interacting neutrons. Subsequently, we have shown that the 
usage of curved spacetime leads to a significant enhancement of the maximum 
mass limit of the ideal neutron star from $0.71$ M$_{\odot}$ to $0.83$ 
M$_{\odot}$. However, such an increase alone cannot explain the existences of 
high mass neutron stars. Nevertheless, the authors have recently shown that the 
consideration of even simple $\sigma-\omega$ model of nuclear interactions in 
the curved spacetime of a non-rotating neutron star, leads to the mass limits 
that are more than  $2 M_{\odot}$ \cite{hossain2021higher}. One can easily 
incorporate such a phenomenological model of nuclear interaction even in the 
curved spacetime of a slowly rotating neutron star as considered here, in order 
to obtain higher mass limits.

\begin{figure}
\includegraphics[height = 7cm, width = 9cm]{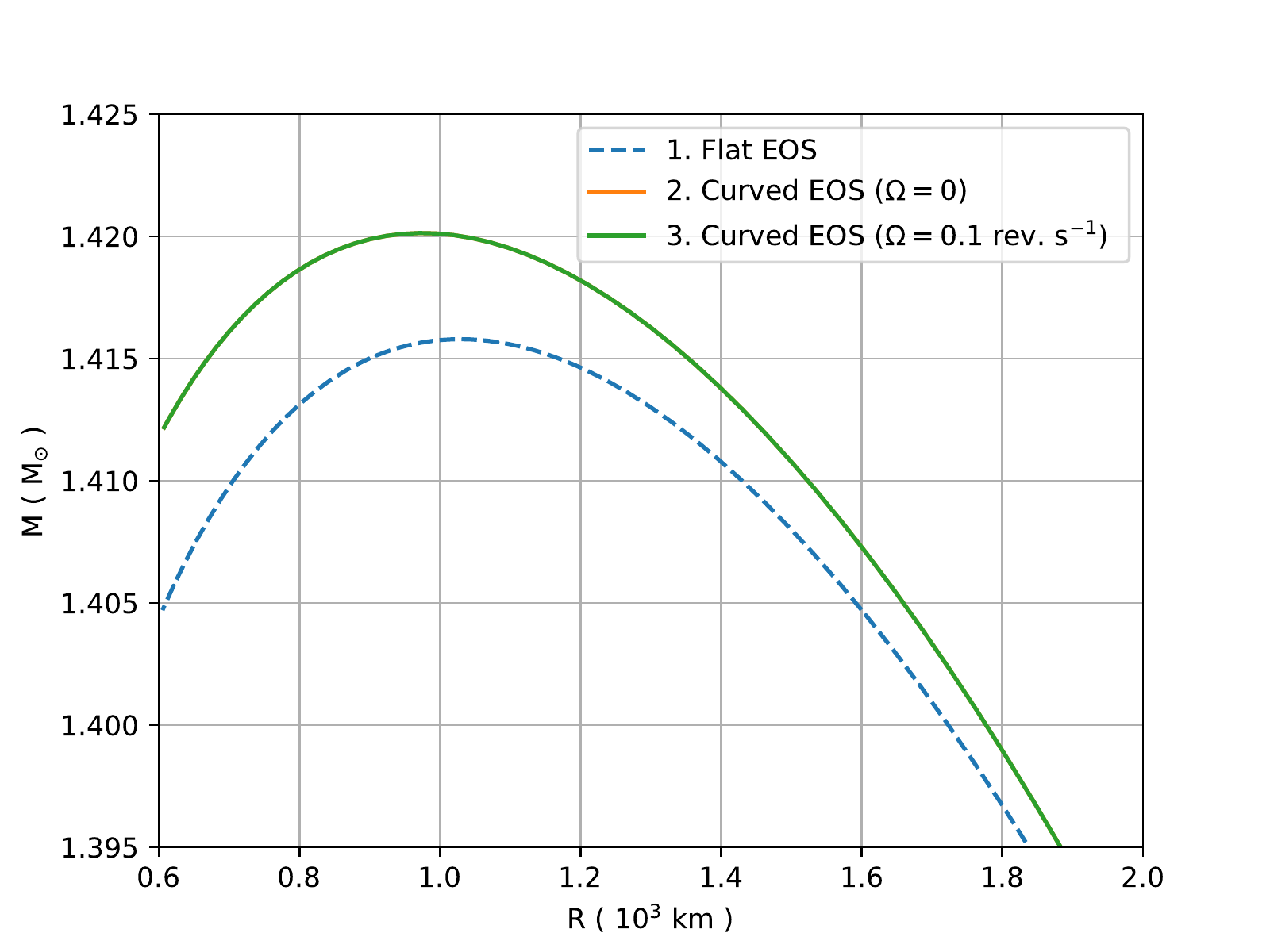}
\caption{The mass-radius relations for a slowly-rotating white dwarf star 
whose degenerate core is made of an ensemble of non-interacting electrons. The 
energy density due to the nuclei is taken to be $\rho_N = (A/Z) m_u n_e 
e^{\Phi}$ where $n_e$, $m_u$, $A$ and $Z$ are the electron number density, 
atomic mass unit, atomic mass number and atomic number respectively 
\cite{hossain2021equation}. For simplicity, here we have taken $A/Z = 2$. 
Similar to the case of neutron star, usage of the curved EOS leads to a higher 
maximum mass limit compared to the usage of flat EOS. We note that the curves 
$2$ and $3$ with different values of $\Omega$ are nearly indistinguishable. In 
other words, the enhancement of the mass limit due to the dragging of inertial 
frames is extremely small in comparison to the time-dilation effect.} 
\label{fig:mr_wd}
\end{figure}

On the other hand, for the white dwarf stars the consideration of 
non-interacting degenerate electrons alone can lead to the maximum mass limit 
which can explain current astrophysical observations. In the FIG. 
\ref{fig:mr_wd}, the mass-radius relations for a slowly-rotating white dwarf 
star are plotted. In particular, the usage of the curved spacetime of a slowly 
rotating white dwarf star leads to an enhancement of the Chandrasekhar mass 
limit form $1.416$ M$_{\odot}$ to $1.420$ M$_{\odot}$ . 

We note that for both neutron stars and white dwarfs the effects of curved 
spacetime manifest through the gravitational time-dilation effect and the 
frame-dragging effect. The effect on the mass limit due to the frame-dragging 
however is extremely small in comparison to the time-dilation effect. By using 
the expansion of the EOS (\ref{PressurePM}, \ref{EnergyDensityPM}) as 
$\rho(n,\Phi,\omega) = \rho(n,\Phi,0) + \mathcal{O}((\omega/m)^2)$, subject to 
the constraint (\ref{OmegaEqn}), one can estimate the enhancement $\Delta M$ of 
the mass $M = \int_{0}^{R}4\pi r^2\rho$ as $\Delta M/M \sim (\Omega/m)^2$. We 
have discussed earlier that the ratio $(\Omega/m) \sim 10^{-22}$ for an ideal 
neutron star which is revolving a hundred times per second and its tiny effects 
on the EOS can be seen in the FIG. \ref{fig:pressure-comparison} and FIG. 
\ref{fig:eos_ratio}.

\subsection{Speed of sound and causality}

It can be seen from the FIG. \ref{fig:eos_ratio} that the effects of 
gravitational time dilation and the dragging of inertial frames both lead the 
equation of state to become relatively \emph{stiffer} compared to its flat 
spacetime counterpart. If the frame-dragging effect is turned off \emph{i.e} if 
we set $\omega \to 0$ then the squared speed of sound $c_s^2 \equiv (dP/d\rho)$ 
reduces to the one corresponding to a spherical spacetime which was shown to 
respect the causality \cite{hossain2021equation}. For high neutron number 
densities \emph{i.e.} $(bn_{\pm}) \gg 1$, the pressure becomes $P 
\approx 
 (e^{\Phi} m^4/24\pi^2) [(bn_{+})^{4/3} + (bn_{-})^{4/3}]$ whereas the energy 
density becomes $\rho \approx  (e^{\Phi} m^4/8\pi^2) [(bn_{+})^{4/3} + 
(bn_{-})^{4/3}]$. This in turn implies that at higher densities, irrespective 
of the values $\Phi$ and $\omega$, the squared speed of sound can be expressed 
as
\begin{equation}\label{SpeedOfSound}
c_s^2 = \frac{dP}{d\rho} \simeq \frac{1}{3} ~.
\end{equation}
At lower densities, as it can bee seen from the FIG. \ref{fig:eos_ratio}, the 
equation of state becomes relatively less stiffer. Therefore, the propagation 
speed of sound within the degenerate matter described by the curved EOS, 
respects causality for both higher and lower neutron number densities.

\section{Discussions}

In summary, we have presented a first-principle derivation of the equation of 
state for an ensemble of degenerate fermions by using the curved spacetime of a 
slowly rotating star. The derived equation of state is directly applicable for 
the studies of degenerate stars such as the neutron stars as well the white 
dwarf stars. We have shown that in contrast to the equation of states 
that are computed in a globally flat spacetime and routinely used in the study 
of neutron stars in the literature, the equation of state computed in the 
curved spacetime depends on two key effects of the general theory of 
relativity, namely the gravitational time dilation effect and the frame-dragging 
effect. Further, we have shown that both of these effects lead to a relatively 
stiffer equation of state which in turn enhances the maximum mass limit of the 
neutron stars. We have also shown that for a given central number density the 
usage of curved spacetime in computing matter EOS, rather than a globally flat 
spacetime, leads to a relatively higher angular momentum for these stars.

The effect due to the gravitational time dilation however has much stronger 
impact on stiffening of the equation of state as compared to the impact of the 
frame-dragging effect. Nevertheless, in general relativity, a non-vanishing 
frame-dragging effect is essential to obtain a non-vanishing angular momentum 
for a rotating star. Further, we have shown here that the frame-dragging effect 
leads to a novel mechanism for stiffening of equation of states by breaking the 
spin-degeneracy of fermions. We are aware that one can also have breaking of 
spin-degeneracy under the influence of an external magnetic field. Since most of 
the compact stars such as the neutron stars are known to contain large magnetic 
fields such a symmetry breaking of the spin-degeneracy is expected to be 
much larger. Thus the mechanism as seen here suggests one to explore the effect 
of external magnetic field on stiffening equation of state due to the breaking 
of spin-degeneracy and to study its consequences. In addition, we note that the 
breaking of spin-degeneracy due to the frame-dragging effect also leads to a 
novel mechanism for generation of seed magnetism in spinning astrophysical 
bodies \cite{SeedMagnetism2022}. Finally, we have seen that the effect of 
curved spacetime on the equation of state is quite significant in the case of 
a neutron star. So the results as shown here suggest a re-look at the various 
studies related to the neutron star that are performed using equation of states 
computed in a globally flat spacetime.

\begin{acknowledgments}
SM thanks IISER Kolkata for supporting this work through a doctoral fellowship. 
GMH acknowledges support from the grant no. MTR/2021/000209 of the SERB, 
Government of India.
\end{acknowledgments}

\vfill

\bibliographystyle{apsrev}
%\bibliography{bibtexfile}

\end{document}